\documentclass[iop,twocolumn]{aastex631}

\usepackage{amsmath}
\DeclareMathOperator{\sech}{sech}

\usepackage{gensymb}
\usepackage{hyperref}
\usepackage{graphicx,latexsym}

\newcommand\xone{x_1}
\newcommand\xtwo{x_2}
\newcommand\xthree{x_3}

\newcommand\Msun{\; {M}_{\odot}}

\newcommand\kms{\; {\rm km/s}}
\newcommand\pc{\;{\rm pc}}

\newcommand\kpc{\;{\rm kpc}}

\newcommand\Gyr{\;{\rm Gyr}}

\newcommand\C{{\mathcal C}}

\newcommand\simgt{\lower.5ex\hbox{$\; \buildrel > \over \sim \;$}}
\newcommand\simlt{\lower.5ex\hbox{$\; \buildrel < \over \sim \;$}}
\newcommand{\RNum}[1]{\uppercase\expandafter{\romannumeral #1\relax}}

\newcommand\PSA{$\Omega_p-A_2$}

\newcommand\PS{$\Omega_p$}
\newcommand\LzA{$\Delta L_z-A_2$}
\newcommand\Rratio{$\cal R$}
\newcommand\Rd{$R_d$}
\newcommand\massratio{$M_{\mathrm per}/M_{\mathrm gal}$}

\newcommand\agama{{\sc agama}}
\newcommand\revision[1]{{#1}}

\def\spose#1{\hbox to 0pt{#1\hss}}
\def\dt{\spose{\raise 1.0ex\hbox{\hskip2pt$\mathchar"201$}}}

\shortauthors{Zheng et al.}

\begin{document}

\title{Comparison of bar formation mechanisms I: does a tidally-induced bar rotate slower than an internally-induced bar?}

\title{Comparison of bar formation mechanisms I: does a tidally-induced bar rotate slower than an internally-induced bar?}

\author[0000-0001-7707-5930]{Yirui Zheng}
\affiliation{Department of Astronomy, School of Physics and Astronomy, Shanghai Jiao Tong University, 800 Dongchuan Road, Shanghai 200240, P.R. China}
\affiliation{Key Laboratory for Particle Astrophysics and Cosmology (MOE) / Shanghai Key Laboratory for Particle Physics and Cosmology, Shanghai 200240, P.R. China}

\author[0000-0001-5604-1643]{Juntai Shen}
\correspondingauthor{Juntai Shen}
\email{jtshen@sjtu.edu.cn}
\affiliation{Department of Astronomy, School of Physics and Astronomy, Shanghai Jiao Tong University, 800 Dongchuan Road, Shanghai 200240, P.R. China}
\affiliation{Key Laboratory for Particle Astrophysics and Cosmology (MOE) / Shanghai Key Laboratory for Particle Physics and Cosmology, Shanghai 200240, P.R. China}


\begin{abstract}

Galactic bars can form via the internal bar instability or external tidal perturbations by other galaxies. We systematically compare the properties of bars formed through the two mechanisms with a series of controlled $N$-body simulations that form bars through internal or external mechanisms. 
We create three disk galaxy models with different dynamical ``hotness'' and evolve them in isolation and under flyby interactions.
In the cold and warm disk models, where bars can form spontaneously in isolation, tidally-induced bars are promoted to a more ``advanced'' evolutionary stage.
However, these bars have similar pattern speeds to those formed spontaneously within the same disk.
Bars formed from both mechanisms have similar distributions in pattern speed--bar strength ($\Omega_p-A_2$) space and exhibit comparable ratios of co-rotation radius to bar length (${\cal R}={R_{\mathrm {CR}}}/{R_{\mathrm {bar}}}$). Dynamical analyses suggest that the inner stellar disk loses the same amount of angular momentum, irrespective of the presence or intensity of the perturbation, which possibly explains the resemblance between tidally and spontaneously formed bars.
In the hot disk model, which avoids the internal bar instability in isolation, a bar forms only under perturbations and rotates more slowly than those in the cold and warm disks. Thus, if ``tidally-induced bars'' refer exclusively to those in galaxies that are otherwise stable against bar instability, they indeed rotate slower than internally-induced ones. However, the pattern speed difference is due to the difference in the internal properties of the bar host galaxies, not the different formation mechanisms.

\end{abstract}



\section{Introduction}
\label{sec:intro}
Galaxy bars are common structures in disk galaxies. Roughly half of spiral galaxies in the nearby universe contain bars in optical bands, with this fraction increasing to two-thirds in infrared bands \citep{marinova2007characterizing,menendez2007near, Erwin2018, Lee2019bar}. Bars are thought to play an important role in galaxy evolution, including triggering gas inflow, influencing star formation, and fostering the development of pseudo-bulges \citep{masters2011galaxy, Li2015, Lin2017, Lin2020, Iles2022}. Our Milky Way is also a barred galaxy \citep[e.g.][]{deVaucouleurs1964, Blitz1991}, and the Milky Way's boxy-bulge and gas dynamics are considered directly linked to the bar \citep[e.g.][]{Shen2010, Li2018}. The widespread presence and significant impact on host galaxies make the study of bar formation and evolution a key topic in galaxy dynamics.


Bars in galaxies can form through two primary mechanisms: internal bar instability and external perturbations. Internally, bars can develop spontaneously due to the gravitational instability of the disks \citep{hohl1971numerical,ostriker1973numerical,sellwood2014secular,Lokas2019iso}. Alternatively, the external mechanism suggests that bar formation can be induced by various gravitational perturbations, such as flyby interaction with other galaxies, collisions, mergers, or the tidal influence of galaxy clusters \citep{byrd1986tidal,gauthier2006substructure, MartinezValpuesta2017, Lokas2016,Lokas2019tidal}.

Several studies reported that bars formed due to tidal interactions (tidal bar or tidally-induced bar) rotate more slowly than those that develop in isolation.
\cite{Miwa1998} found that tidally-induced bars rotate slowly, with inner Lindblad resonances (ILRs) \revision{located} near their ends, while spontaneous bars lack ILRs due to their rapid rotation.
\cite{MartinezValpuesta2017} demonstrated that bars triggered by interactions stay in the slow regime for a longer time than those formed via internal instability.
\cite{Lokas2018} also found that bars formed under the influence of a perturbing body are stronger and exhibit lower pattern speeds.
Dwarf satellites orbiting a Milky Way-like central galaxy host slower bars, consistent with the primary galaxy cases mentioned above \citep{Lokas2016, Gajda2017, Gajda2018}.

However, the comparison of the pattern speed of bars formed under the two mechanisms needs to be made more carefully for at least two reasons.
First, if perturbations promote or delay bar formation, tidally-induced bars may not be at the same evolutionary stage as the spontaneous bars, while it is well known that bars slow down during their evolution.
Dynamical friction between the dark matter halo and the stellar bar transfers angular momentum from the bar to the halo, slowing down the bar with modest growth \citep{Weinberg1985Apr, Debattista2000}.
Trapping disk stars onto the elongated orbit also decreases the rotation rate but grows the bar at the same time \citep{Athanassoula2003}.
Another reason is that the bar pattern speed is strongly relevant to the host galaxy properties.
For instance, the bar pattern speed is found to be correlated with the stellar mass \citep{GarmaOehmichen2020}, the bulge fraction \citep{Kataria2019}, and disc circular speed \citep{GarmaOehmichen2022}.
Therefore, it is more meaningful to control related galaxy properties when comparing the pattern speed of tidally-induced bars with spontaneous ones.
A more systematic investigation is desired to understand the different rotation rates between the bars formed through the two mechanisms.

To mitigate the influence of galaxy properties on the bar's pattern speed, it is common to use the dimensionless parameter \Rratio$=R_{\mathrm {CR}}/R_{\mathrm {bar}}$, where $R_{\mathrm {CR}}$ is the co-rotation radius and $R_{\mathrm {bar}}$ is the bar length.
Analyses of the shapes of dust lanes predict the \Rratio\ parameter to fall within a range of $1.2\pm0.2$ \citep{Athanassoula1992}.
A widely used threshold of 1.4 is employed to differentiate between fast and slow bars \citep[e.g.][]{Debattista2000, Corsini2008, Fathi2009, Aguerri2015}.
Fast bars are defined by $1.0\leq$ $\cal{R}$ $ \leq 1.4$, with 1.0 being the lower limit as bars in this range can become unstable and dissolve since there are no bar-supporting orbits beyond the co-rotation radius \citep{Contopoulos1989}.
Slow bars have $\cal{R}$ $ > 1.4$.
Observational studies predominantly found fast bars in the real universe \citep{Aguerri2015, Guo2019}, yet many simulations, including cosmological ones \citep{Algorry2017, Peschken2019, Roshan2021}, tidal interactions \citep{Lokas2016, Gajda2017, Gajda2018}, produce bars in the slow regime.
It is not fully understood why simulated bars are slower than observed ones.

State-of-the-art cosmological simulations have proven powerful in many aspects of galaxy formation and evolution, yet controlled simulations are still necessary to study the mechanisms of bar formation. 
The concerns are twofold. 
First, results from different cosmological simulations are inconsistent with observations and each other. For instance, bars are suppressed in low-mass galaxies in the IllustrisTNG100 simulation \citep{zhao2020}, while the EAGLE simulation produces the highest bar fraction in a similar low-mass range \citep{cavanagh2022evolution}. 
Moreover, bars in cosmological simulations are generally less prominent than those in the real universe, often appearing too short \citep{zhao2020, Frankel2022} and/or rotating too slowly \citep{Roshan2021}. 
The second concern stems from the difficulty in effectively controlling the intensity of interactions in cosmological simulations. 
Unfortunately, the shapes and structures of individual galaxies are influenced by both their environment and internal baryonic physics, making their morphologies often unpredictable \citep{Zhou2020}. 
To systematically study bar properties under different formation mechanisms, it is necessary to conduct controlled simulations using a series of galaxy models and controlled perturbation scenarios.

The structure of the paper is as follows. In Section \ref{sec:sims}, we detail the galaxy models and the simulations conducted, including both isolated and flyby interaction scenarios. In Section \ref{sec:results}, we present the simulation outcomes, comparing specifically the pattern speed of bars formed through internal instability versus external perturbation mechanisms. In Section \ref{sec:discussion}, we analyze the implications of our results within the context of galaxy evolution. Finally, in Section \ref{sec:summary}, we conclude the paper with a summary of our key findings.


\section{Simulations}
\label{sec:sims}

\subsection{Galaxy models}
\label{sec:models}

To compare the bar properties formed via internal instability and external perturbation, we generate three different galaxy models and allow them to evolve in isolation and under flyby interactions.
Each model has a stellar disk embedded in a live dark matter halo, devoid of a bulge component. 
These stellar disks vary in stability, categorized as cold, warm, and hot disks. 
In isolation, the cold and warm disks spontaneously form bars, although the warm disk requires more time to do so. 
In contrast, the hot disk fails to form a bar after evolving for 6\Gyr. However, when subjected to flyby interactions, all galaxy models are capable of bar formation. The tidal perturbations can promote bar formation in the cold disk, although they are not the sole determining factor.
In the warm disk, tidal perturbations can significantly accelerate the bar formation process, resulting in earlier and faster bar development compared to spontaneous formation. 
For the hot disk, tidal perturbations are the only way to form a bar.

Our galaxy models are constructed using \agama\ \citep{AGAMA2019}, an action-based software library tailored for a wide range of applications in stellar dynamics. Each galaxy model exhibits the same density profile, comprising an exponential quasi-isothermal disk and a Hernquist dark matter (DM) halo. The density profile of the stellar disk is given by:
\begin{equation}
    \rho_* = \frac{M_*}{4\pi R_d^2 h_z} \exp(-R/R_d) \sech^2(z/h_z).
\label{eq:rho_star}
\end{equation}
where $M_*$ represents the mass of the stellar disk, $R_d$ is the scale length, and $h_z$ is the scale height. Across our models, the stellar disks are consistent in mass, with $M_* = 3.6 \times 10^{10} \Msun$, and scale length, with $R_d = 2 \kpc$. The scale heights are uniform $h_z = 0.4 \kpc$, resulting in a disk thickness ratio of $h_z/R_d = 0.2$.

The dark matter (DM) halos in our galaxy models follow the \revision{truncated} Hernquist profile as below:
\begin{equation}
  \rho_{\rm DM} = \frac{M_{\rm halo}}{2\pi} \frac{a}{r(r+a)^3} \times \exp \left[-(r/r_{\rm cut})^2 \right].
\end{equation}
This equation consists of two parts: the first part is the standard Hernquist profile, where $M_{\rm halo}$ is the total mass of the halo and $a$ is the scale radius of the profile; the second part is a cutoff term that reduces the density at large radii, with $r_{\rm cut}$ being the cutoff radius. For all our DM halos we select $M_{\rm halo} = 3.6 \times 10^{11} \Msun$, $a = 13.7 \kpc$, and $r_{\rm cut} = 114 \kpc$. The total mass of stars and dark matter is $M_{\rm tot} = 4.0 \times 10^{11} \Msun$.
 



\agama\ establishes the velocity distribution of the stellar disk using an action-based distribution function (DF). 
We refer readers to \citet{AGAMA2019} for a detailed description. Here we only highlight the radial velocity dispersion, the key physical quantity related to the stabilities of the disks. The radial velocity dispersion profile of the stellar disk is given by:
\begin{equation}
    \sigma_R(R) = \sigma_{R,0} \exp(-R/R_{\sigma, R}),
\label{eq:sigmaR}
\end{equation}
where $\sigma_{R,0}$ represents the velocity dispersion at the galaxy center, and $R_{\sigma, R}$ is the scale length of the profile. 
Following the recommendation of \citet{AGAMA2019}, we set $R_{\sigma, R} = 2\;R_d$, resulting in $R_{\sigma, R} = 4 \kpc$ for our galaxy models. 
To differentiate between the cold, warm, and hot disks, we assign the central velocity dispersions of $\sigma_{R,0} = 73 \kms$, $124 \kms$, and $226 \kms$, respectively.

\begin{figure}
    \centering
    \includegraphics[width=\columnwidth]{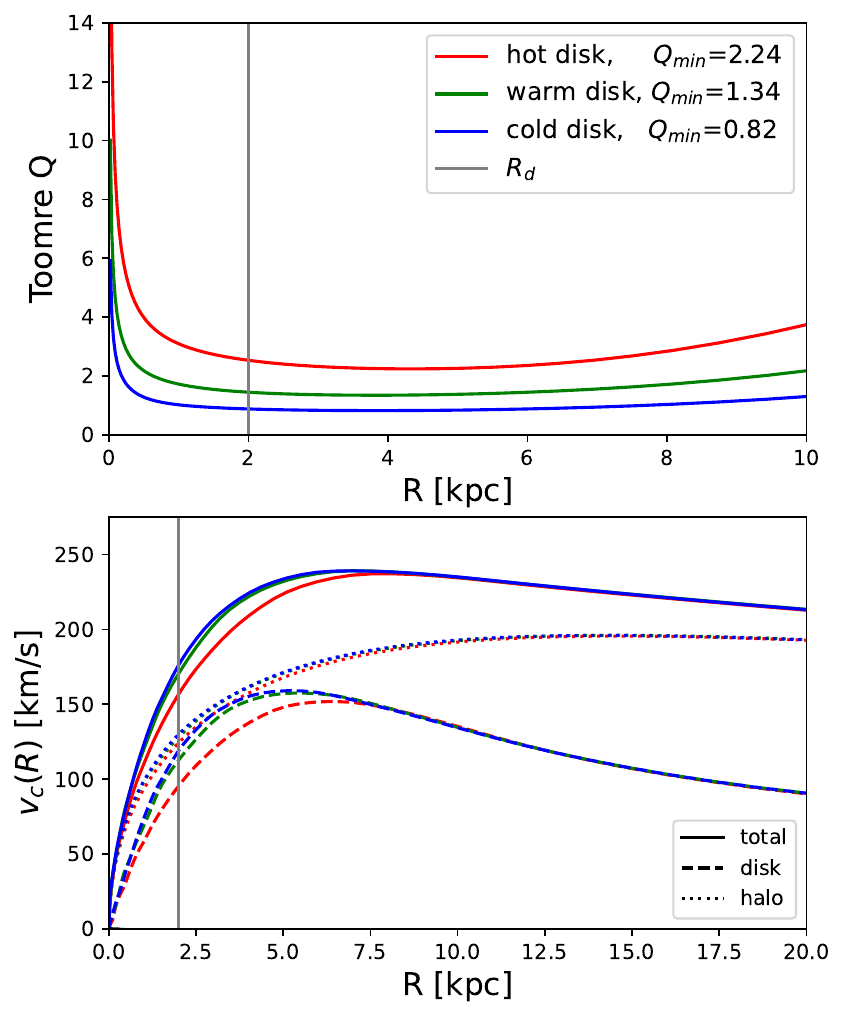}
    \caption{Toomre $Q$ profiles ({\it upper panel}) and rotation curves ({\it lower panel}) of the stellar disks in different galaxy models. The solid lines represent the values for the cold (blue), warm (green), and hot (red) models, with the minimum $Q$ values for each model noted in the legend. 
    The dotted lines in the lower panel represent the halo's contribution to the rotation curve, while the dashed lines indicate the contribution from the stellar disk.
    The vertical grey line marks the location of the stellar disk's scale length ($R_d=2\kpc$).
    }
    \label{fig:toomreQ}
\end{figure}

The stability of a disk is often quantified by the Toomre $Q$ parameter, which is expressed as:
\begin{equation}
    Q = \frac{\kappa \sigma_R}{3.36 G \Sigma},
\label{eq:toomreQ}
\end{equation}
where $\kappa$ is the epicycle frequency, $\sigma_R$ is the radial velocity dispersion, $G$ is the gravitational constant, and $\Sigma$ is the surface density of the disk. 
The Toomre $Q$ profiles of our galaxy models are plotted in the upper panel of \autoref{fig:toomreQ}. 
The Toomre $Q$ curves exhibit similarities across the three disk models due to the identical density and velocity dispersion profile shapes of the stellar disks. 
With different $\sigma_{R,0}$, the cold, warm, and hot disks have minimum \revision{Toomre stability  values $Q_{\mathrm{min}}$ }  of 0.82, 1.34, and 2.24, respectively.
\revision{
\citet{Jang2023} found that bar formation requires
\begin{equation}
(Q_{\text{min}}/1.2)^2 + (\text{CMC}/0.05)^2\lesssim 1,
\end{equation}
where CMC represents the central mass concentration, which is 0 in our models. 
Substituting the $Q_{\mathrm{min}}$ values, we obtain results of 0.47, 1.25, and 3.48 for the cold, warm, and hot disks, respectively. 
This indicates that the cold disk can form a bar, the warm disk is marginally stable, and the hot disk is stable against bar instability.
}

In the lower panel of \autoref{fig:toomreQ}, we present 
\revision{the circular speed curves $v_c(R)$ }
for different models along with the contribution from the halo (dotted lines) and that from the stellar disk (dashed lines). These models show considerable similarities, especially in the outer regions of the galaxies, which is a natural consequence of employing identical parameters for the stellar disk and dark matter halo density profiles. We notice that the rotation curve of the hot disk is slightly lower than those of the cold and warm disks.
\agama\ performs iterations to achieve the desired density and velocity dispersion profiles when combining the stellar disk with the dark matter halo. 
The significantly higher velocity dispersion in the hot disk leads to a slightly greater deviation from the initially given parameters, resulting in a lower \revision{circular speed} in the inner region compared to the cold and warm disks.


\subsection{Setup of the flyby interaction}
\label{subsec:flyby}

Utilizing these galaxy models, we initiate a series of flyby interaction simulations. For simplicity, the perturber is a pure Hernquist dark matter halo made of live particles. The mass ratio of the perturber to the galaxy, \massratio, is set to 1/1, 1/3, and 1/10, respectively, thereby generating strong, moderate, and weak tidal perturbations to the galaxy.

We configure the galaxy and the perturber along an orbit inspired by the setup in \citet[see her Figure 1 for a schematic view]{Lokas2018}. 
For all our flyby simulations, the separation between the galaxy and the perturber is fixed at 280\kpc\ in the $x$-direction and 14\kpc\ in the $y$-direction. 
The relative velocity is 560\kms\ in the $x$-direction and 0 in the $y$-direction. 
$v_z$ of both the galaxy and the perturber are 0, ensuring that the centroids of both objects remain on the $x-y$ plane. 
In this setup, the closest approach occurs around $t = 0.5$\Gyr\ for all flyby simulations.
The pericenter distances are approximately 7, 9, and 11\kpc\ for mass ratios of 1/1, 1/3, and 1/10, respectively.
For each mass ratio, we vary the inclination angle between the stellar disk plane and the orbit plane of the dark matter perturber to induce prograde ($0^\circ$), perpendicular ($90^\circ$), and retrograde ($180^\circ$) flyby interactions. Consequently, we conduct 9 flyby simulations for each disk model.

\subsection{Simulation details}
\label{subsec:simdetails}
We perform all $N$-body simulations with \texttt{GADGET-4} \citep{Springel2005,Springel2021}. 
Each simulation is evolved for 6\Gyr.

We assign 0.5 million particles to the stellar disk and 1 million particles to the dark matter halo, resulting in a particle mass of $m_* = 7.2\times10^4 \Msun$ for the stellar disk and $m_{\rm DM} = 3.6 \times 10^5 \Msun$ for the DM halo. 
The gravitational softening length is set to 23\pc\ for the stars and 57\pc\ for the DM particles.

As we are not concerned with the evolution of the perturber, we configure it as a pure dark matter halo with reduced resolution. 
In the equal-mass interaction, where the perturber has a mass of $4 \times 10^{11} \Msun$, it is composed of $2 \times 10^5$ DM particles, yielding a particle mass of $m_{\rm DM} = 2 \times 10^6 \Msun$. The softening length is set to 100\pc.
For the unequal-mass interactions, the number of perturber particles is set to 1/3 and 1/10 of the equal-mass case to maintain the same mass resolution.


\section{Results}
\label{sec:results}

\subsection{Bar strength and pattern speed}
\label{subsec:barpro}

\begin{figure*}
    \centering
    \includegraphics[width=\textwidth]{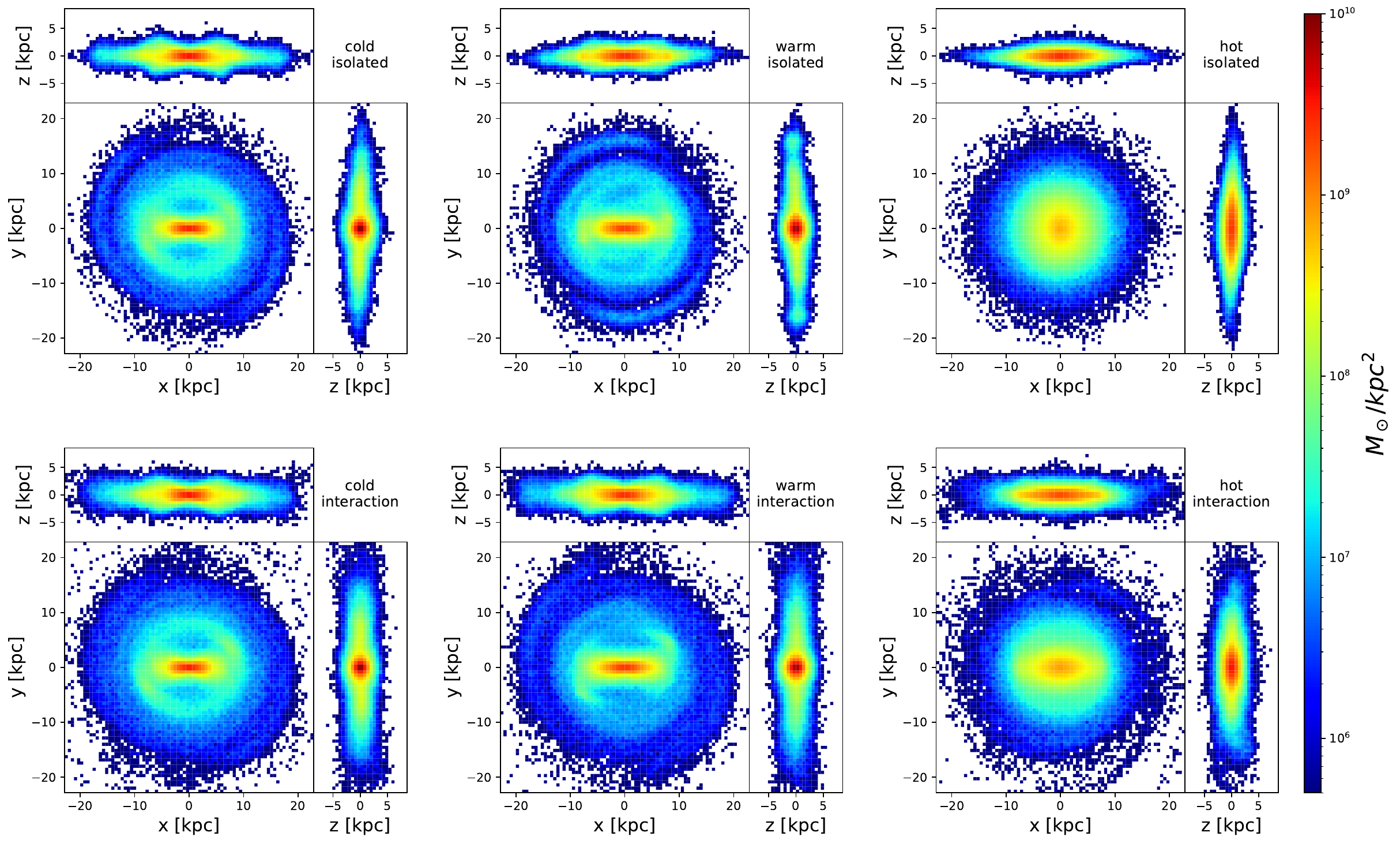}
    \caption{The stellar surface density of the cold, warm, and hot disks at the end of isolated simulations ({\it upper row}) and flyby interactions ({\it lower row}, \massratio$=1/3$, perpendicular orbit). In isolated cases, the cold and warm disks spontaneously form bars, while the hot disk remains unbarred throughout. In the flyby interactions, all disks form bars, with the hot disk hosting a relatively weak bar.}
\label{fig:gal_plot}
\end{figure*}

\begin{figure}
\centering
\includegraphics[width=\columnwidth]{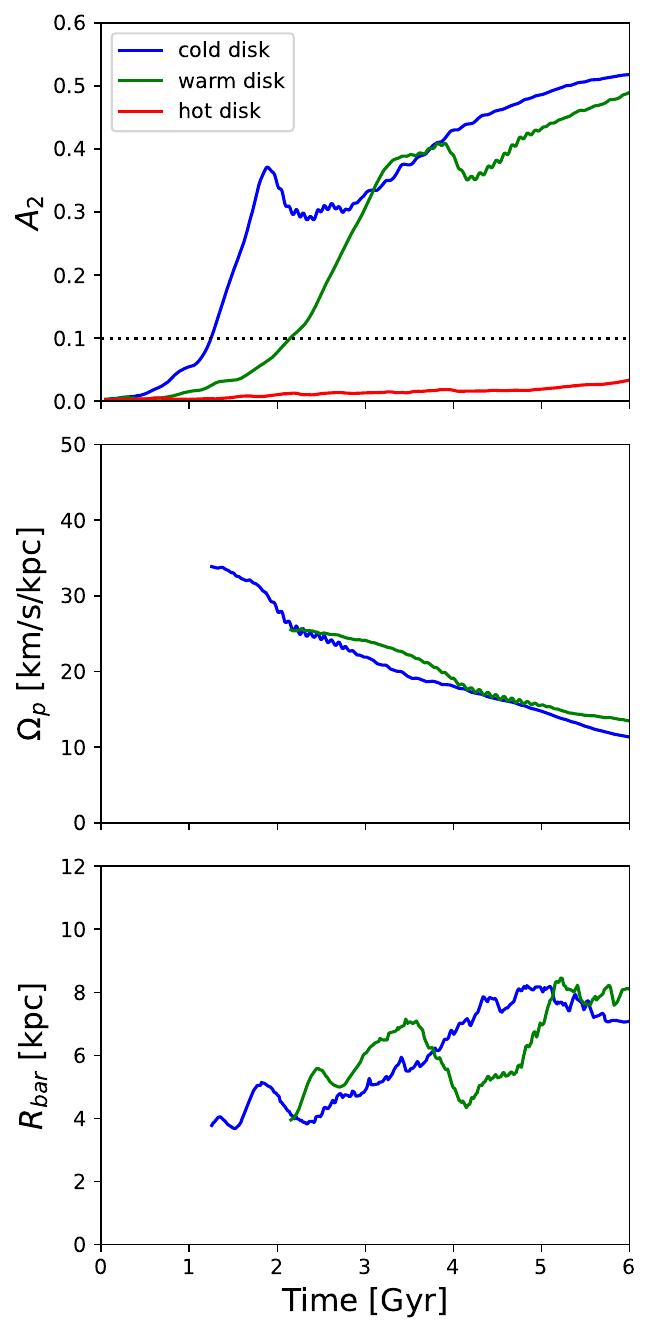}
\caption{Bar strength $A_2$ ({\it upper panel}), pattern speed \PS\ ({\it middle panel}) and bar length $R_{\mathrm {bar}}$ ({\it lower panel}) for isolated cold, warm, and hot disks. The cold and warm disks spontaneously form a bar, while the hot disk does not develop a bar within 6\Gyr. The dotted line marks $A_2 = 0.1$, representing the minimum bar strength required for calculating \PS\ and $R_{\mathrm {bar}}$.
}
\label{fig:a2ps_iso}
\end{figure}

\begin{figure*}
\centering
\includegraphics[width=0.85\textwidth]{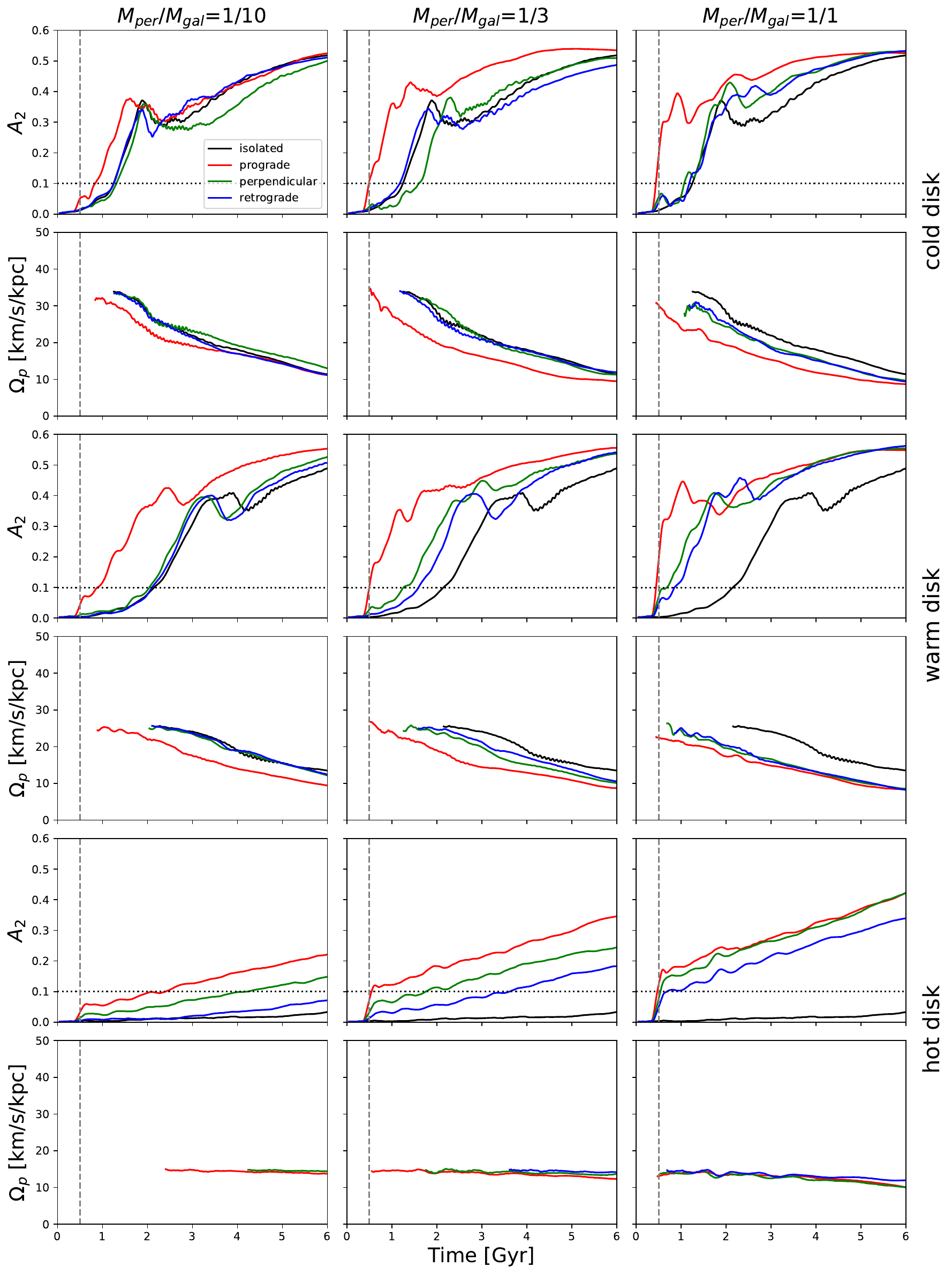}
\caption{
Bar strength $A_2$ (odd rows) and pattern speed \PS (even rows) for flyby interactions. From left to right, the mass ratios of the pure dark matter perturber to the galaxy (\massratio) are 1/10, 1/3, and 1/1, respectively. The first two rows show the results for the cold disk model; the third and fourth rows represent the warm disk; and the last two rows are for the hot disk. Within each panel, the outcomes of tidal interactions are presented in colored lines, while the isolated cases are also plotted in black for comparison.
The grey vertical dashed line marks $t=0.5\Gyr$, representing the intended time of closest approach for the perturber.  }
\label{fig:a2ps_flyby}
\end{figure*}

We plot the stellar surface density of the three models at the end of the isolated simulations ($t=6\Gyr$) in the upper row of \autoref{fig:gal_plot}.
The cold and the warm disk form bars in their isolated simulations while the hot disk remains axisymmetric throughout.
The lower row of \autoref{fig:gal_plot} shows the result of the flyby interactions with a perturber mass ratio of 1/3 and a perpendicular orbit.
In these cases, all disks form bars, with the hot disk hosting a relatively weak bar.
To quantify the bar properties, we measure the bar strength, pattern speed, and length.

Various methods have been developed to quantify bar strength in the literature \citep{Combes1981, Athanassoula2003}. 
The Fourier amplitude of the $m=2$ mode, $A_2$, is relatively more widely used, especially in the papers cited in our introduction and discussion sections. 
Therefore, we employ $A_2$ to quantify bar strength in our paper to ensure comparability with previous works.

We first compute the sine and cosine coefficients of the $m=2$ mode of the stellar disk using the following equations:
\begin{subequations}
\begin{align}
    a_2 &= \frac{ \sum_{i=1}^{N} m_i \cos(2\phi_i)} {\sum_{i=1}^{N} m_i}, \\
    b_2 &= \frac{ \sum_{i=1}^{N} m_i \sin(2\phi_i)} {\sum_{i=1}^{N} m_i},
\end{align}
\end{subequations}
where $m_i$ and $\phi_i$ are the mass and the azimuthal angle of the $i$-th particle, respectively. The summation is performed over the stellar particles within a cylindrical region with $R \leq 4\;R_d$, i.e. $R \leq 8 \kpc$ for our galaxy models. 
The limit of $4\;R_d$ is selected as it approximately corresponds to the maximum radius of the bar in our simulations.
We also \revision{test different} radius of $3\;R_d$ \revision{and $6\;R_d$, finding} that all conclusions are unchanged.

The bar strength $A_2$ is then calculated as:
\begin{equation}
    A_2 = \sqrt{a_2^2 + b_2^2},
\end{equation}
while the position angle of the bar $\phi_{\mathrm {bar}}$ is :
\begin{equation}
   \phi_{\mathrm {bar}} = \frac{1}{2} \arctan({b_2}/{a_2}).
\end{equation}
The pattern speed \PS\ is then defined as the difference in position angle over time, i.e.,
\begin{equation}
    \Omega_p = {\Delta \phi_{\mathrm {bar}}}/{\Delta t}.
\end{equation}

We plot the bar strength $A_2$ (upper panel) and the pattern speed \PS\ (middle panel) for the isolated simulations in \autoref{fig:a2ps_iso}. 
In pattern speed measurement, we only consider bars with $A_2 \geq 0.1$ to avoid noise, especially during the early stages of bar formation when $\phi_{\mathrm {bar}}$ is not well-defined.
We also display the bar length $R_{\mathrm{bar}}$ for the bars in the cold and warm disks in the lower panel of \autoref{fig:a2ps_iso}. The measurement and analyses of $R_{\mathrm{bar}}$ are detailed in subsection \ref{subsec:rratio}.

\autoref{fig:a2ps_iso} illustrates that the cold and warm disks can spontaneously form bars, with the warm disk taking longer to achieve the same bar strength. 
The hot disk does not form a bar within 6\Gyr. Its consistently low $A_2$ value shows the hot disk remains axisymmetric throughout the simulation. 
These three models represent real galaxies with varying stabilities against bar formation in isolation, and they are employed to investigate the impact of tidal perturbations on bar formation.

\autoref{fig:a2ps_flyby} compares the bar strength $A_2$ and the pattern speed \PS\ of flyby simulations with those of the isolated ones.
In some simulations, there are small, transient spurs or spiral arms attached at the ends of the bars, which can lead to fluctuations in the bar strength and the pattern speed. 
To mitigate the influence of these spurs or spiral arms, we apply a moving average with a time window of 0.2\Gyr\ to the bar strength and the pattern speed.

The first two rows show the results of the cold disk model. When compared to the isolated cases, a prograde-moving perturber can promote the development of bar structures, resulting in earlier and more rapid bar formation. 
The promoted bars appear to rotate at a slower pace than their isolated counterparts (red lines in the second row of \autoref{fig:a2ps_flyby}). 
In contrast, a perturber moving perpendicular or retrograde has a significantly reduced impact on bar formation promotion. 
In the perpendicular case with a 1/3 mass ratio, the bar emerges even slightly later than in the isolated scenario.
The pattern speeds in the perpendicular and retrograde cases are comparable to those in the isolated scenarios. 
These findings suggest that bar formation in the cold disk is primarily driven by internal disk instability, with only extreme external perturbations capable of accelerating the appearance of bars.


In the third and fourth rows of \autoref{fig:a2ps_flyby}, the results for the warm disk model are displayed. The warm disk also forms a bar spontaneously, but it does so on a longer timescale than the cold disk model. 
Perturbations prove to be significantly more effective in promoting bar formation in the warm disk, except in the very weak cases where the perturber has a 1/10 mass ratio and follows a perpendicular or retrograde orbit. 
In these simulations, bar formation is a result of a combination of internal disk instability and external perturbation.
The tidally-affected bars exhibit lower pattern speeds compared to their spontaneously formed counterparts. The greater the perturbation is, the earlier the bar formation occurs and the slower the bar rotates.

\revision{
    The cold and warm disk models are susceptible to the bar instability in isolation.
    While stronger tidal forces accelerate the emergence of bars, the fundamental mechanism driving bar formation in these disks may still be their internal bar instability. 
    \cite{Moetazedian2017} proposed that tidal forces introduce additional perturbations that spread inward radially and are then swing-amplified.
    Hence, the bars formed under tidal interactions in these disks are not purely tidal-induced, but rather a combination of internal and external mechanisms.   
    To be exact, these bars are ``bars under the influence of tidal forces''.
    For convenience, we still use the term ``tidally-induced bars'' loosely throughout the paper.
}

We present the results of the hot disk model in the final two rows of \autoref{fig:a2ps_flyby}. The hot disk does not spontaneously form a bar in isolation, but the tidal perturbation does gradually induce a bar in most instances. 
A visual inspection of the density plots reveals that the tidal arms contribute to a rapid increase in $A_2$ shortly after the pericenter passage of the perturber at $t=0.5\Gyr$. 
These tidal arms are unstable and rapidly dissipate, with $A_2$ signals then dominated by the gradually emerging stellar bars. 
It is important to note that these bars differ from those in the cold and warm disks. In the hot disk, the bar is nearly extended across the entire disk and exhibits smaller ellipticities, and is like an oval disk (see an example in \autoref{fig:gal_plot}).
The pattern speeds of the bars in the hot disk are highly similar, and they remain almost constant throughout the simulation, with only slight decreases over time.

\revision{
    We note that the pattern speeds of the tidal bars in the hot disk models are determined by the intrinsic properties of the disk instead of the perturber's angular velocity. 
    The angular velocities of the perturbers at the pericenter are approximately 70, 63, and 56 ${\mathrm{km/s/kpc}}$ for mass ratios of 1/1, 1/3, and 1/10, respectively. These values significantly exceed the pattern speeds of the tidal bars in the hot disk models. Furthermore, simulations conducted with increased perturber velocities revealed that the pattern speeds of the tidal bars remain largely unchanged. }


The bars in the cold and warm disks gradually decelerate after their formation, which is anticipated and commonly observed in the literature \citep{Weinberg1985Apr, Debattista2000, Athanassoula2003}. 
This deceleration is attributed to the capture of disk stars \revision{to the outer parts of the bar} and the dynamical friction between the bar and the halo, with stronger bars experiencing more pronounced deceleration. 
Tidally-induced bars typically form earlier and are at a more advanced evolutionary stage compared to spontaneously formed bars in the same disk. Consequently, they undergo a longer duration of slowing down at a fixed simulation time. 
The relatively lower position of the colored lines in the even rows of \autoref{fig:a2ps_flyby} should not be interpreted solely as indicating slower rotation of tidal bars. The lower pattern speed values can also be a direct consequence of the more advanced evolutionary stage of the tidally-induced bars.

\subsection{Pattern speed--bar strength (\PSA) space}
\label{subsec:psa}
\begin{figure}
\centering
\includegraphics[width=\columnwidth]{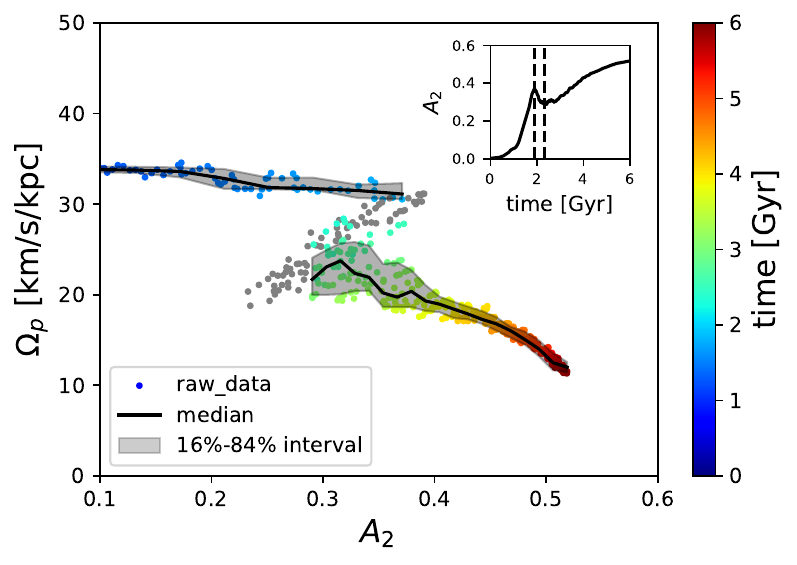}
\caption{The \PSA\ plot of the cold disk evolved in isolation. The scatter points show the raw, unsmoothed data, color-coded by simulation time.
The data points corresponding to the buckling stage are depicted in grey to prevent distraction (see the text for further details).
We then divide the data during the formation and the secular growth stages into several bins based on $A_2$, plotting the median \PS\ in each bin with a solid line. The shaded region around the median line indicates the 16th and 84th percentiles. 
In the inset axes, the solid line shows the evolution of the bar strength $A_2$ as a function of time, while the dashed lines indicate the end of the bar formation stage and the beginning of the secular growth stage following buckling. \revision{By checking the vertical asymmetry, we confirm that the decrease in $A_2$ is indeed due to the buckling of the bar, rather than other mechanisms}}
\label{fig:psa_eg}
\end{figure}

\begin{figure*}
    \centering
    \includegraphics[width=\textwidth]{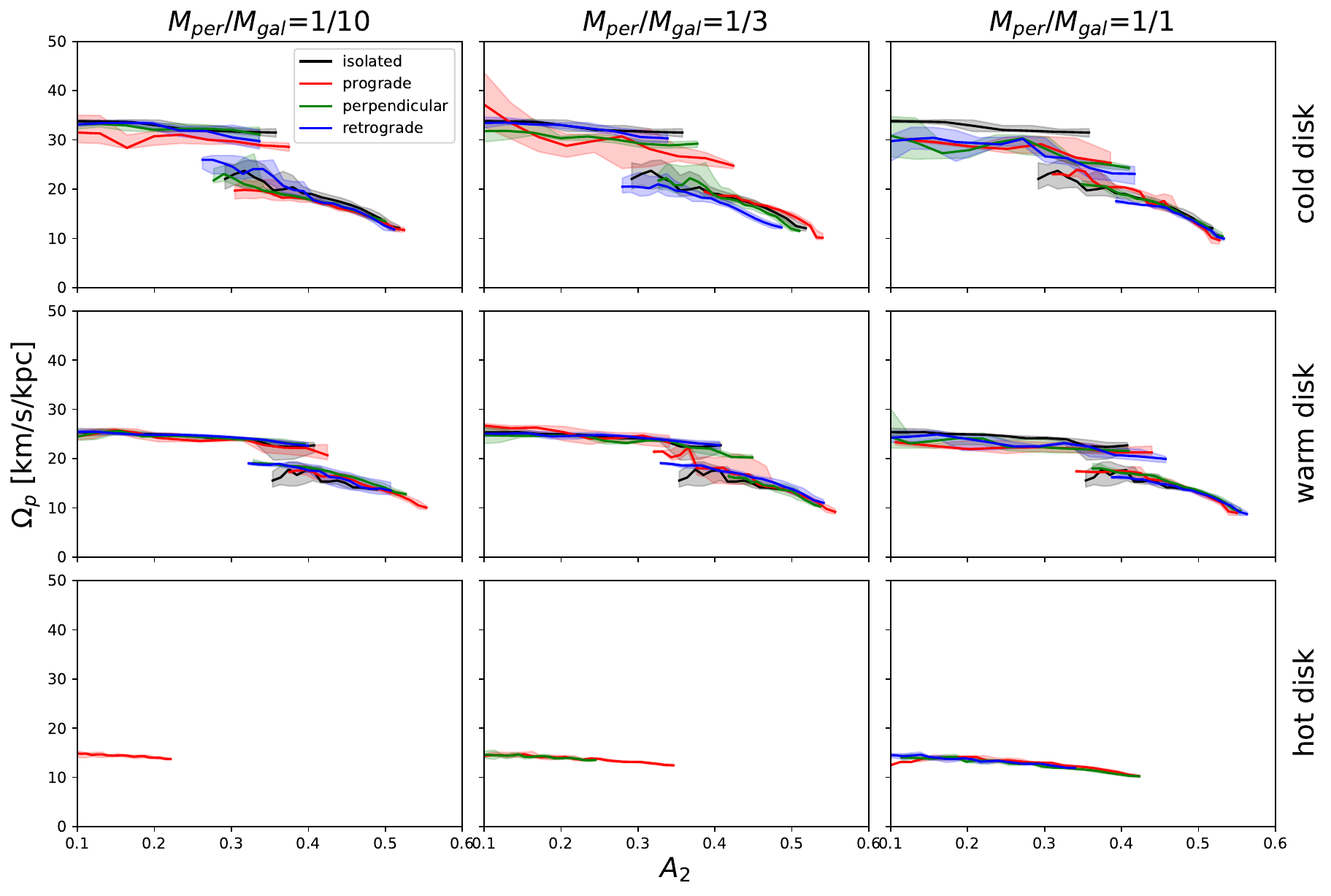}
    \caption{The \PSA\ space of all simulations. 
    The panel layout and color scheme mirror those of \autoref{fig:a2ps_flyby}. For the cold and warm disk models, the tidally-induced bars fall onto the same \PSA\ distribution as the spontaneously formed bars in the same disk. As for the hot disk, all tidally-induced bars exhibit the same \PSA\ distribution.
    }
    \label{fig:psa_all}
    \end{figure*}

One must consider the different stages of bar evolution when comparing the pattern speeds of tidal and spontaneous bars. 
A straightforward method to illustrate the bar evolutionary stage is through the bar strength $A_2$. 
In this section, we will present pattern speed--bar strength (\PSA) space for different simulations and investigate whether tidal bars exhibit the same \PSA\ distribution as their spontaneously formed counterparts in the same disk in isolation.

\autoref{fig:psa_eg} illustrates the \PSA\ space for the cold disk in isolation. 
The raw data from each snapshot are divided into three stages: 
1) the bar formation stage, where the smoothed $A_2$ increases until reaching its maximum value; 
2) the buckling stage, where $A_2$ decreases from its maximum to a local minimum;
3) the secular bar growth stage following buckling, where $A_2$ increases again. 
The buckling stages vary in strength and duration among the simulations, making direct comparisons less informative. Hence, we focus on the formation stage and the secular growth stage of the bar.

For each growth stage, we divide the data into several bins of $A_2$
and calculate the median and the 16th and 84th percentiles of the pattern speed \PS\ in each bin.
These metrics nicely track the trend of \PS\ against $A_2$ in each stage, as shown in \autoref{fig:psa_eg}.
We observe anti-correlation between the pattern speed and the bar strength in these two stages, which is consistent with previous studies \citep[e.g.][]{Athanassoula2003, Berentzen2004}.

We carry out the same reduction on the \PSA\ distribution for all simulations and show the results in \autoref{fig:psa_all}. 
Taking the \PSA\ distribution of the same disk in isolation as the reference (black line and grey shaded region), we find that the tidally-induced bars in the cold and warm disks fall onto \revision{simliar} \PSA\ distribution \revision{to that of} the internally-induced bar in the same disk.

The conclusion above is solid at the secular growth stage following buckling for all mass ratios and all orbit inclinations. 
During the bar formation stage, however, we find that the tidally-induced bars in the cold or warm disk may exhibit slightly lower \PS\ than their spontaneously formed counterparts in some cases of very strong perturbation, such as the 1/1 mass ratio and the prograde orbit. 
In these simulations, strong tidal spiral arms arise due to the perturbation, which can substantially affect the calculation of $A_2$, thus rendering the Fourier analysis less reliable for characterizing bar strength. 
Additionally, the rapid bar growth in these cases makes it challenging to compute the \PS\ accurately 
in the bar formation stage.

We highlight the \massratio$=1/3$ interaction series in the warm disk model as the best demonstration of the \PSA\ distribution. 
As shown in the middle panel of the 4th row of \autoref{fig:a2ps_flyby}, the tidally-induced bars consistently exhibit lower pattern speeds than their spontaneously formed counterparts at fixed simulation time. 
However, in the middle panel of the second row of \autoref{fig:psa_all}, the \PSA\ distribution for the tidally-induced bars in the \massratio$=1/3$ interaction series aligns with that of the spontaneously formed bars in the same disk in isolation.
This finding suggests that when considering bar strength, the tidally-induced bars rotate at the same rate as the spontaneously formed bars within the same disk in isolation.

In the bottom row of \autoref{fig:psa_all}, the \PSA\ distribution for the hot disk is consistent across various flyby simulations.  
For a hot disk that is stable against bar formation in isolation, the tidally-induced bars rotate at the same speed irrespective of the perturbation's intensity, provided that the bar strength is taken into account.

\subsection{\LzA \ space}
\label{subsec:lza}
\begin{figure*}
\centering
\includegraphics[width=\textwidth]{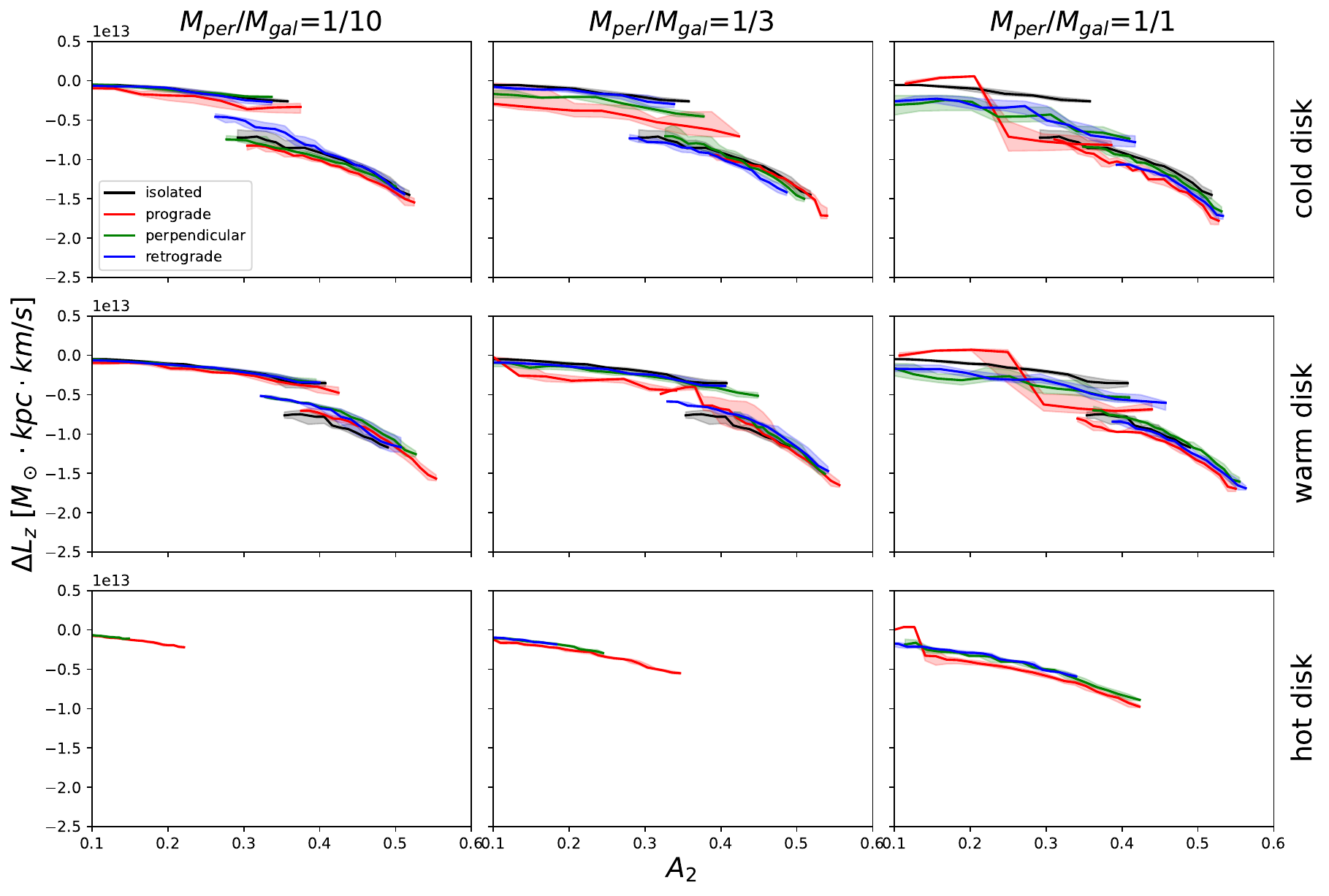}
\caption{The \LzA\ space of all simulations. The panel layout and color scheme mirror those of \autoref{fig:a2ps_flyby}. The y-axis represents the change in angular momentum $\Delta L_z$ of the stellar disk within 4\;\Rd. Our results indicate that the inner stellar disk loses an equivalent amount of angular momentum regardless of the presence or absence of perturbation.
}
\label{fig:lza_all}
\end{figure*}

We attempt to interpret the results in the \PSA\ space through a kinematic analysis of the stellar disk. 
We compute the angular momentum, $L_z$, of the stellar disk within 4\;\Rd, which is the same range utilized in the Fourier analysis described in \autoref{subsec:barpro}. 
We then compare this value with the initial condition to determine the amount of angular momentum lost, $\Delta L_z$: 
\begin{equation}
    \Delta L_z = L_z (<4\;R_d) - L_{z,ini}(<4\;R_d),
\end{equation}
where $L_z(<4\;R_d)$ represents the angular momentum of the disk within 4\;\Rd\ , and $L_{z,ini}(<4\;R_d)$ is the corresponding value of the initial condition.
We also test the choice of 3\;\Rd\ \revision{or 6\;\Rd ,} and find that a \revision{different} radius does not alter the conclusions in this section.
The results are plotted against the bar strength $A_2$, and we apply the same reduction method as in the \PSA\ space to visualize the \LzA\ space for all simulations in \autoref{fig:lza_all}.

For the same disk, all simulations converge to the same \LzA\ distribution, which indicates that the inner stellar disk loses the same amount of angular momentum regardless of the perturbation's presence or intensity. 
This consistency is particularly pronounced during the secular bar growth stage following buckling, while it is slightly less evident during the bar formation stage, due to the presence of strong spiral arms and the rapid growth of bars in some simulations. 
In these cases where the distribution deviates, we find that the tidally-induced bars lose more angular momentum than the spontaneously formed bars in the same disk, and simultaneously, the tidal bars have a lower \PS\ than the spontaneous bars.

The similarity between the \PSA\ and \LzA\ distribution highlights the critical role of angular momentum in deciding the pattern speed of bars.
It also corroborates our conclusion that tidally-induced bars are at a more advanced evolutionary stage compared to spontaneously formed bars in the same disk, with the relatively lower \PS\ of the tidal bars being attributed to the longer duration of slowing down at a fixed simulation time.

For the hot model, a consistent \LzA\ distribution is observed across all simulations, further reinforcing that angular momentum is a critical factor of the pattern speed of bars.

\subsection{\Rratio \ ratio}
\label{subsec:rratio}
\begin{figure*}
\centering
\includegraphics[width=0.96\textwidth]{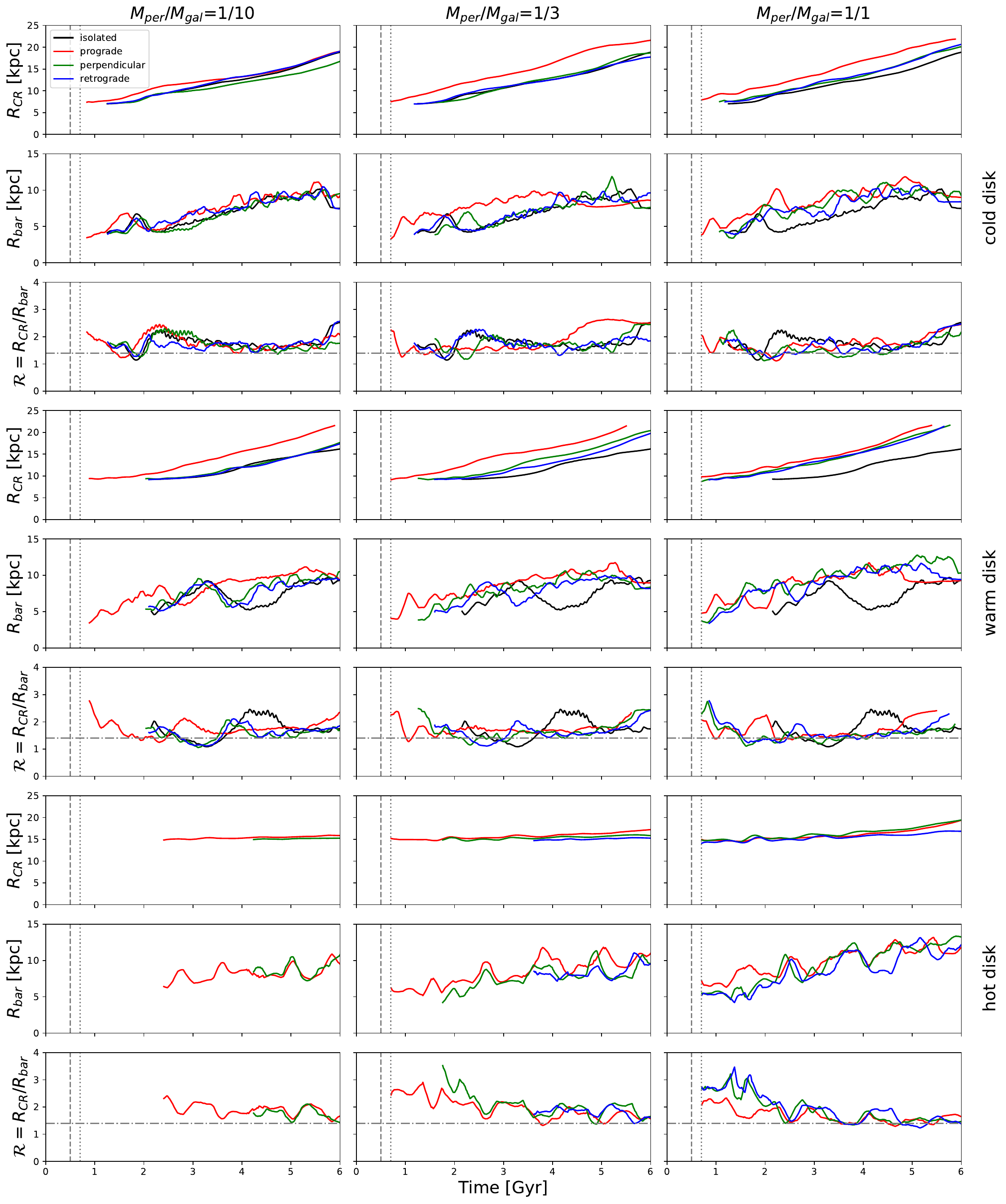}
\caption{The \Rratio\ ratio of all simulations. The panel layout and color scheme mimic those in \autoref{fig:a2ps_flyby}. The y-axis represent the co-rotation radius $R_{\mathrm {CR}}$, the bar  length $R_{\mathrm {bar}}$ and their ratio $\cal{R} $$= R_{\mathrm {CR}}/R_{\mathrm {bar}}$. 
The division between fast and slow bars (\Rratio $=1.4$) is plotted with a horizontal dash-dotted line for reference.
We show the data after the bar strength reaches $A_2=0.1$ to avoid noise in the \Rratio\ ratio calculation. 
Data before $t=0.7\Gyr$(the dotted line) are not shown as the galaxy has not yet settled down after the closest approach of the perturber ($t=0.5\Gyr$, the dashed line).
We stop computing $R_{\mathrm {CR}}$ when $R_{\mathrm {CR}}\gtrsim 10\;R_d$, where there are not enough stellar particles to reliably trace stellar acceleration in \texttt{GADGET-4}.
Our results indicate that the tidally-induced bars exhibit the same \Rratio\ ratio as the spontaneously formed bars within the same disk.}
\label{fig:rratio}
\end{figure*}

An alternative method to quantify the bar rotation speed is the \Rratio\ ratio, defined as the ratio of the co-rotation radius to the bar length, $\cal{R}$$ = R_{\mathrm {CR}}/R_{\mathrm {bar}}$ \citep{Debattista1998, Debattista2000}. In observational studies, \Rratio $=1.4 $ is typically considered the upper limit for fast bars, with \Rratio $> 1.4$ characterizing slow bars. It is not uncommon to observe a higher \Rratio\ ratio in simulations compared to observations, which is also the case in our study. Consequently, we solely compare the \Rratio\ ratio of the tidal and spontaneous bars within the same disk, disregarding the absolute value of the \Rratio\ ratio.

The co-rotation radius is defined as the radius at which the pattern speed of the bar matches the circular frequency of the disk, that is, $\Omega_p = \Omega_c$. 
We first compute $\Omega_c$ using the equation:
\begin{equation}
    \Omega_c (R) = v_c(R)/R=\sqrt{F_R(R) / R},
\end{equation}
where $v_c(R)$ is the circular speed at radius $R$, and
$F_R(R)$ is the azimuthally-averaged radial force. 

The measurement of the bar length is more sophisticated. Numerous methods have been applied to quantify the bar length, although the outcomes are not always in agreement with one another \citep{Ghosh2024}. 
In recent work, \cite{Kataria2022} employed five different methods to measure the bar length and found that the length determined by the variation of the bar phase angle is close to the average length derived from these methods and matches the visual estimate.

Following the approach outlined in their study, 
we divide the stellar disk into a series of annuli and compute the phase angle of the $m=2$ mode of the stellar disk within each annulus. 
The bar length is then defined as the radius at which the phase angle of the $m=2$ mode deviates from a nearly constant value by 5$^\circ$. 
Detailed examples of the bar length measurement process are provided in \autoref{app:barlength}. We further validate this definition by randomly selecting more snapshots from our simulations and confirming that it aligns with the visual assessment of the density plots in the majority of cases.
However, we admit that the bar length may not be well-defined in the early stages of bar formation, particularly when the bar growth is rapid in the strong interaction cases. 
Additionally, in the later stages, the method may underestimate the bar length when the bar is distorted (e.g., during buckling) or overestimate it when small transient spiral arms at the ends of the bars happen to vanish during the measurement. 
To mitigate these issues, we average the bar length over a time range of 0.2\Gyr\ to reduce the noise.

The co-rotation radius $R_{\mathrm{CR}}$, bar length $R_{\mathrm{bar}}$, and \Rratio\ ratio of all simulations are shown in \autoref{fig:rratio}. 
To minimize noise in the \Rratio\ ratio calculation, we only display the data after the bar strength attains $A_2=0.1$. 
Data before $t=0.7\Gyr$ are omitted since the galaxy has not fully stabilized following the perturber's closest approach at $t=0.5\Gyr$. This exclusion is necessary to avoid unreliable measurements of the co-rotation radius and, consequently, the \Rratio\ ratio.

In the first three rows of \autoref{fig:rratio}, we present the results for the cold disk model. The co-rotation radius $R_{\mathrm{CR}}$ increases gradually as the bar slows down. The bar length $R_{\mathrm{bar}}$ fluctuates in tandem with the bar strength $A_2$ (see the first row of \autoref{fig:a2ps_flyby}). For instance, when the bar in the isolated simulation undergoes buckling around $t=2\Gyr$, a decrease in bar length is observed, which is also reflected in the \Rratio\ ratio. 
At any given time, the tidally-induced bars feature larger co-rotation radii compared to the internally-induced one in the same disk, as the \revision{tidal ones have} a lower pattern speeds. Concurrently, the bar length of the tidal bars is greater than that of the spontaneous bar. 
Thus, bars formed in the cold disk through different mechanisms display the same \Rratio\ ratio range of $1 < $\Rratio $< 2$, as shown in the third row of \autoref{fig:rratio}.
This result supports our conclusion that tidally-induced bars are at a more advanced evolutionary stage than spontaneously formed bars within the same disk. The relatively lower \PS\ of the tidal bars does not necessarily indicate that they rotate slower; 
the tidal bars are as fast as their counterparts in isolation when the bar length is taken into account.

We show the results for the warm disk in the 4th, 5th, and 6th row of \autoref{fig:rratio}. 
In this model, bar formation is primarily driven by external perturbation when the perturbation is significant. 
We again observe that the \Rratio\ ratio of the tidally-induced bars is similar to that of the spontaneously formed bar within this disk model.

The last three rows of \autoref{fig:rratio} display the result of the hot disk model.
For the hot disk model, the conclusion remains that the tidally-induced bars rotate at the same speed regardless of the perturbation's strength.

When comparing different models,
the bars in the hot disk have a larger co-rotation radius $R_{\mathrm {CR}}$ and longer bar length $R_{\mathrm {bar}}$ compared to the cold and warm disks.
The \Rratio\ ratio of the tidally-induced bars in the hot disk has a high initial value greater than 2, unlike the cold and warm disks.
However, as the bar length increases, the \Rratio\ ratio decreases and eventually converges to a value of around 1.5, comparable to that observed in the cold and warm disks.


\section{Discussion}
\label{sec:discussion}

\subsection{Do tidal bars rotate slower?}
\label{subsec:comparison}

We find that tidally-induced bars rotate at the same speed as their internally-induced counterparts within the same disk when the bar strength (\autoref{fig:psa_all}) and length (\autoref{fig:rratio}) are taken into account. 
However, our conclusions do not necessarily contradict previous studies that conclude tidal bars rotate more slowly. 
In \cite{Miwa1998}, \cite{Lokas2016}, \cite{Gajda2017,Gajda2018}, and \cite{Lokas2018}, 
the initial disk galaxy models in the tidal bar simulations were stable enough to inhibit global bisymmetric instability, 
which is distinct from the models that spontaneously develop bars.
These studies actually compare bars formed in different galaxies, rather than bars in the same galaxy but formed through different mechanisms.

\begin{figure}
    \centering
    \includegraphics[width=\columnwidth]{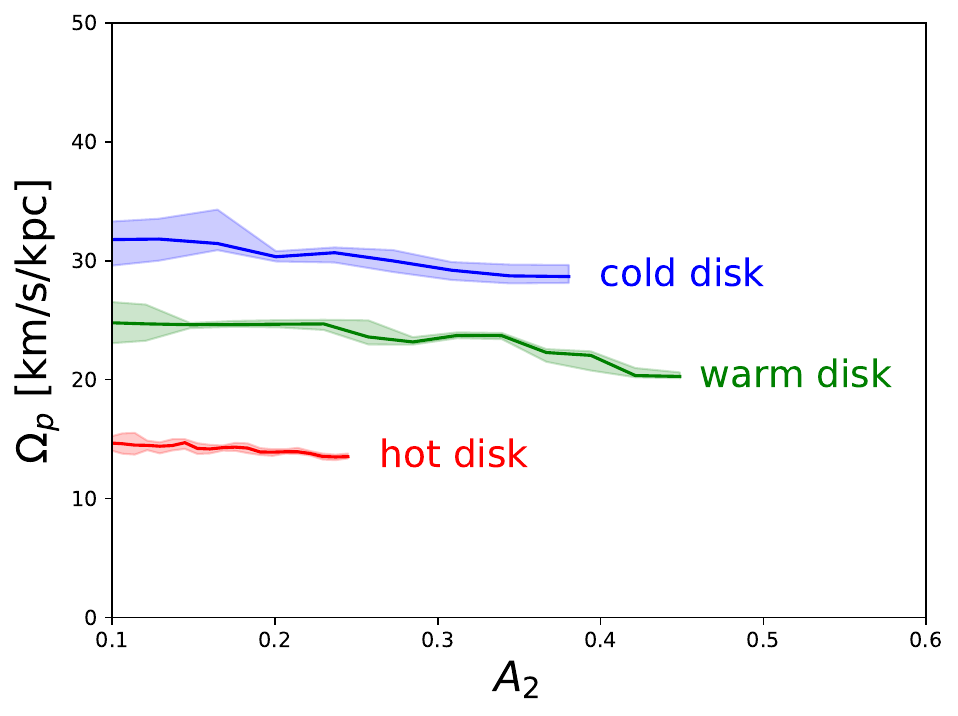}
    \caption{Comparison of the \PSA\ distribution of different galaxy models. 
    For each model, we select the 1/3 mass ratio with a perpendicular orbit, representing an intermediate perturbation in our simulations. Only the bar formation stage is shown.
    The tidally-induced bar in our hot disk rotates slower than the ones in the cold and warm disks.
    }
    \label{fig:psa_diff}
\end{figure}

To compare bars across our models, we select one simulation from each disk model and plot their \PSA\ distribution in \autoref{fig:psa_diff}. 
We chose the 1/3 mass ratio with a perpendicular orbit, representing an intermediate perturbation in our simulations.
We note that any selection should yield consistent conclusions as the bars within the same disk exhibit the same \PSA\ distribution.
Only the bar formation stage is shown.
We exclude the secular growth stage following buckling to avoid clutter. Additionally, the bar in the hot disk model lacks a secular growth stage as it does not buckle.

\autoref{fig:psa_diff} illustrates that the tidally-induced bar in our hot disk model rotates slower than the ones in the cold and warm disk models, thus also slower than the spontaneous bars in these two cooler models since all bars in the same disk share the same \PSA\ distribution.
If one restricts the term ``tidally-induced bars'' only to bars in galaxies that are stable against bar formation in isolation, our ``tidally-induced bars'' rotate slower than spontaneous ones, which aligns with previous studies.
However, we also notice that the bar in the warm disk rotates slower than the one in the cold disk.
We argue that the difference in rotation speed is not attributed to the presence of tidal perturbation but the internal nature of the galaxies hosting the bars, which is the different velocity dispersion in our cases. 
This is also the case in \cite{MartinezValpuesta2017}, where the tidal bars in the stable model rotate more slowly than the spontaneously formed bars in the unstable galaxy.

\cite{MartinezValpuesta2017} also concluded that tidally affected bars rotate more slowly than spontaneous bars in the same cold models, but this difference is only pronounced when the strongest perturbation is considered (see their Figure 2). 
In their strongest interaction case, the tidal bars are slightly weaker and shorter than the spontaneously formed bars, which contrasts with our findings that tidal bars are promoted and strengthened by the perturbation when evolution time is held the same. 
The disparity may stem from the distinct manner in which the perturbation is realized: they employ an impulse approximation, whereas we utilize a live dark matter halo. 

\revision{The majority of our flyby interactions promote bar formation, which can be attributed to the fact that tidal perturbations dominate the internally-driven bar modes in the cold and warm disk models as a result of early pericenter times and the presence of massive perturbers. 
Several studies, including \cite{Moetazedian2017}, \cite{Pettitt2018}, and \cite{Zana2018}, have demonstrated that tidal forces do not always promote bar formation and can sometimes delay it. 
This delay may be caused by destructive interference between tidal perturbations and internally-driven bar modes. 
In the perpendicular encounter of the cold disk with a 1/3 mass ratio perturber, we observe a slight delay in bar formation compared to the isolated case. 
The tidal bar in this delayed case also rotates at a similar speed as the internally-induced in the same disk once the bar strength/length is considered, as illustrated in Figures \ref{fig:psa_all} and \ref{fig:rratio}. 
If the pericenter time were postponed (e.g., to $2-3$\;\Gyr), we might observe more instances of delayed bar formation. Nonetheless, the delayed tidal bars would still rotate at the same speed as the spontaneous bars in the same disk, once the differences in the evolutionary stages are properly considered.}

Our results are largely consistent with \cite{Berentzen2004}, who also found that tidal bars within the same disk share the same \PSA\ and \LzA\ distribution. However, while \citeauthor{Berentzen2004} interpreted their results as an indication that tidal bars rotate more slowly, the temporal evolution of the tidal bars was not thoroughly considered. If one examines a fixed time, the tidal bars are indeed slower than the spontaneously formed bars within the same disk, as observed in our \autoref{fig:a2ps_flyby}. Nevertheless, the tidal bars are in an advanced evolutionary stage compared to the spontaneous bars in the same disk. Considering the tendency of bars to slow down after their formation, the apparent disparity in rotation speeds is not a result of the different formation scenarios but rather the different evolutionary stages of the bars.

Our findings suggest that the pattern speed of bars is primarily determined by internal galaxy properties, irrespective of the presence or intensity of tidal perturbation, aligning with the results in \cite{Salo1991}.  
\cite{Gerin1990} also found a bar
with the same pattern speed in the perturbed and unperturbed cases, implying that the perturber does not impose its own pattern speed.
However, \cite{Gerin1990} noted that perturbation on an {\bf already barred disk} can alter the bar strength pattern speed to about 10\%.
The precise value of the bar pattern speed is contingent upon the loss of angular momentum from the inner stellar disk, and tidal perturbation does not alter the angular momentum of the inner disk in the bar region when the bar strength is accounted for (\autoref{fig:lza_all}).

Astronomers can only observe a single snapshot of a galaxy's evolution, not its entire history. It is limited to differentiate between the various evolutionary stages of bars.
Consequently, the disparity in bar rotation speed is not a reliable criterion for distinguishing between tidally-induced and spontaneously formed bars. 
The results of \cite{Berentzen2004} and our work support that tidally-induced bars are comparable to those formed in isolated discs, 
making it challenging (if not impossible) to differentiate them from spontaneously formed bars based on their overall properties, such as pattern speed or bar strength.

A potential method to distinguish tidally-induced bars from spontaneously formed ones is to examine the detailed structure of the bars. 
In our work, the stellar disk initially forms tidal arms due to the perturbation, which then develop into a bar. 
Tidal bars tend to grow more rapidly than spontaneously formed bars. These differences may manifest in distinct shapes and orbit families within the bars.
\revision{Though \autoref{fig:gal_plot} does not show a clear distinction between the tidal and spontaneous bars, a visual assessment of the stellar density contours reveals that bars formed by strong tidal perturbations tend to have a ``slightly thinner waist'' along the bar minor axis. A more detailed examination of the bar's shape and the associated orbit families is required to corroborate these distinctions further.}

Another approach is to investigate the detailed kinematic properties of the bars. 
\cite{Iles2024} found that host disks of tidally-induced bars exhibit larger stellar migration overall, with populations of inner disc stars displaced to large radii and below the disc plane. 
We propose that the detailed velocity dispersion profile and the angular momentum exchange mechanism of the bars and the remaining stellar disks may also differ between tidally-induced and spontaneously formed bars, warranting further examination. 

The long-term objective of this project is to develop reliable methods for differentiating tidally-induced bars from spontaneously formed ones.


\subsection{Effect of the inclination of the perturber orbit}
\label{subsec:inclination}

\cite{MartinezValpuesta2017} found a negligible difference between prograde and retrograde perturbations on the formation of bars, whereas \cite{Lang2014} and \cite{Lokas2018} found that prograde perturbations are more effective in promoting bar formation and generating stronger bars. 
Our results align with the latter works, as shown in \autoref{fig:a2ps_flyby}, where bars in prograde simulations occur earlier and are more pronounced compared to perpendicular and retrograde simulations.

\cite{Lokas2018} proposed that the tidal force exerted by the perturber acts on stars in the prograde disk for a longer duration compared to those in the retrograde disk, which cannot be adequately captured by the impulse approximation employed in \cite{MartinezValpuesta2017}. 
We concur with this explanation and further propose that the halo spin of the primary galaxy also influences the perturbation's impact.
Although the primary galaxy's halo is initially set without net angular momentum, the perturbation can induce the halo to align with the direction of the perturbed. 
At the same time, isolated simulations have proved the important role of inner halo angular momentum in bar formation: bars are initiated earlier in prograde halos, whereas their formation is delayed in counter-rotating halos \citep{Kataria2022}.

In perpendicular perturbation simulations, we observe a minimal difference between the perpendicular and retrograde perturbations in the cold and warm disk models, \revision{except for} a slight delay in bar formation in the 1/3 interaction of the cold disk model. 
However, perpendicular perturbations appear to be more effective in promoting bar formation in the hot disk model. 
We align with the explanation provided in \cite{Lokas2018} and propose that the tidal interaction time on individual stars in the perpendicular perturbation is slightly longer than that in the retrograde perturbation but shorter than in the prograde perturbation. 
The halo spin of the galaxy may also contribute to the perturbation's effect, although this influence is not as pronounced as that of the prograde perturbation.

The different influence on bar formation suggests that the strength of the perturbation is substantially affected by the inclination of the perturber's orbit, even when other parameters are held the same.
\cite{Elmegreen1991} introduced a dimensionless tidal strength parameter $S$ to quantify the perturbation intensity:
\begin{equation}
    S = \frac{M_{\mathrm{per}}}{M_{\mathrm{gal}}} \left( \frac{R_{\mathrm{gal}}}{d} \right)^3 \frac{\Delta T}{T},
\end{equation}
where $M_{\mathrm{per}}$ and $M_{\mathrm{gal}}$ denote the masses of the perturber and the galaxy, respectively. $R_{\mathrm{gal}}$ represents the galaxy size, and $d$ is the distance of closest approach. $\Delta T$ is the time taken by the perturber to orbit 1 radian at closest approach, while $T$ is the time for stars at $R_{\mathrm{gal}}$ to orbit 1 radian around the galactic center.
$S$ is interpreted as the ratio of the angular momentum imparted to an outer star by the perturber to its original orbit momentum.
Under such an interpretation, we argue that the current definition applies only to prograde cases. 
To quantify the perturbation's strength in more general terms, it is necessary to revise the $S$ parameter to incorporate the inclination of the perturber's orbit. 
\revision{The revision of the $S$ parameter and further analysis on how $S$ impacts the bar properties} 
will be an ancillary objective in this series of work.


\section{Summary}
\label{sec:summary}

To systematically investigate the pattern speed of tidal bars, we generate three pure disk galaxy models with varying stabilities against bar formation by adjusting the radial velocity dispersion (\autoref{fig:toomreQ}). 
These galaxies are designated as cold, warm, and hot. 
The cold and warm disk models spontaneously form bars, although the warm disk requires a longer time. The hot disk model does not develop a bar within 6\Gyr\ of isolated evolution (\autoref{fig:a2ps_iso}).

These models are then perturbed by a live dark matter halo with varying mass ratios and orbit inclinations. 
The bar strength and pattern speed are calculated to study the impact of the perturbation on bar formation and properties (\autoref{fig:a2ps_flyby}). 
In the cold disk, bar formation is primarily driven by internal disk instability, whereas only extreme external perturbation can lead to an earlier onset of bars. 
In the warm disk, bar formation results from a combination of internal instability and external perturbation, and we observe that the more intense the tidal interaction, the earlier the bar formation occurs. 
In the hot disk, tidal perturbation is the only mechanism to trigger bar formation, and the tidally-induced bars share a very similar pattern speed.

Given that tidal bars are typically in a more advanced evolutionary stage than their spontaneously formed counterparts within the same disk, 
we plot the \PSA\ space for a more meaningful comparison between the tidal and spontaneous bars (\autoref{fig:psa_eg} and \autoref{fig:psa_all}).
We find that the tidally-induced bars and spontaneously formed bars converge to the same \PSA\ distribution in the cold and warm disk models, suggesting that the tidal bars rotate at the same speed as the spontaneously formed bars within the same disk when the bar strength is considered. 
Further analysis of angular momentum indicates that the tidally-induced bars lose the same amount of angular momentum as the spontaneously formed bars in the same disk (\autoref{fig:lza_all}), which likely explains the similar pattern speeds of the tidal and spontaneous bars.
For the hot disk that avoids internal bar instability, we hold the same conclusion that the tidally-induced bars rotate at the same speed and the inner disk loses the same amount of angular momentum regardless of the perturbation's intensity.

We also compute the bar length and the co-rotation radius to study the \Rratio\ ratio of the bars (\autoref{fig:rratio}). No discernible difference is observed in the \Rratio\ ratio between the tidal and spontaneous bars within the same disk. This result indicates that the tidal bars have the same angular speed as the spontaneously formed bars within the same disk when the bar length is taken into account.

We subsequently discuss our findings in the context of literature that reported that tidal bars rotate more slowly than spontaneously formed ones. 
For bars within the same galaxy, we argue that the seemingly slower pattern speed of tidal bars in these studies is due to their different evolutionary stages, not their formation mechanisms. 
When comparing bars across different galaxies, tidally-induced bars indeed rotate slower than spontaneous ones, if ``tidal bars” refer exclusively to those in galaxies that are stable against isolated bar formation (\autoref{fig:psa_diff}). 
However, the lower rotation speed of these ``tidal bars'' is not due to the presence of external tidal perturbation but rather to the internal nature of the galaxies hosting the bars.

To the best of our knowledge, this work is the first instance where the external formation mechanism is firmly established as not being the cause of the observed lower pattern speed in tidally-induced bars. 
The differences in the pattern speeds of bars are more appropriately linked to the distinct evolutionary stages of the bars and the intrinsic properties of the host galaxies, rather than being solely attributed to the presence of tidal perturbations.


Our results indicate that bars formed through different scenarios cannot be readily distinguished by their rotation rates. 
More detailed research on bar structure and dynamical properties is necessary to 
discern tidally-induced bars from spontaneously formed ones.



\software{
    {\sc agama}\citep{AGAMA2019},
    \texttt{GADGET-4} \citep{Springel2005,Springel2021}. 
    NumPy \citep{2020NumPy-Array},
    SciPy \citep{2020SciPy-NMeth},
    Matplotlib \citep{4160265},
    Jupyter Notebook \citep{Kluyver2016jupyter}
}

\section*{Acknowledgements}
We thank the anonymous referee for suggestions that helped to improve the presentation of the paper. 
We thank Zhi Li and Sandeep Kumar Kataria for their valuable insights on simulations and analysis. We also thank Xufen Wu and Rui Guo for their helpful discussions.
The research presented here is partially supported by the National Key R\&D Program of China under grant No. 2018YFA0404501; by the National Natural Science Foundation of China under grant Nos.  12025302, 11773052, 11761131016; by the ``111'' Project of the Ministry of Education of China under grant No. B20019; and by the Chinese Space Station Telescope project. J.S. acknowledges support from the {\it Newton Advanced Fellowship} awarded by the Royal Society and the Newton Fund. 
This work made use of the Gravity Supercomputer at the Department of Astronomy, Shanghai Jiao Tong University.



\appendix
\restartappendixnumbering
\section{Bar length measurement}
\label{app:barlength}

\begin{figure}
	\includegraphics[width=1\columnwidth]{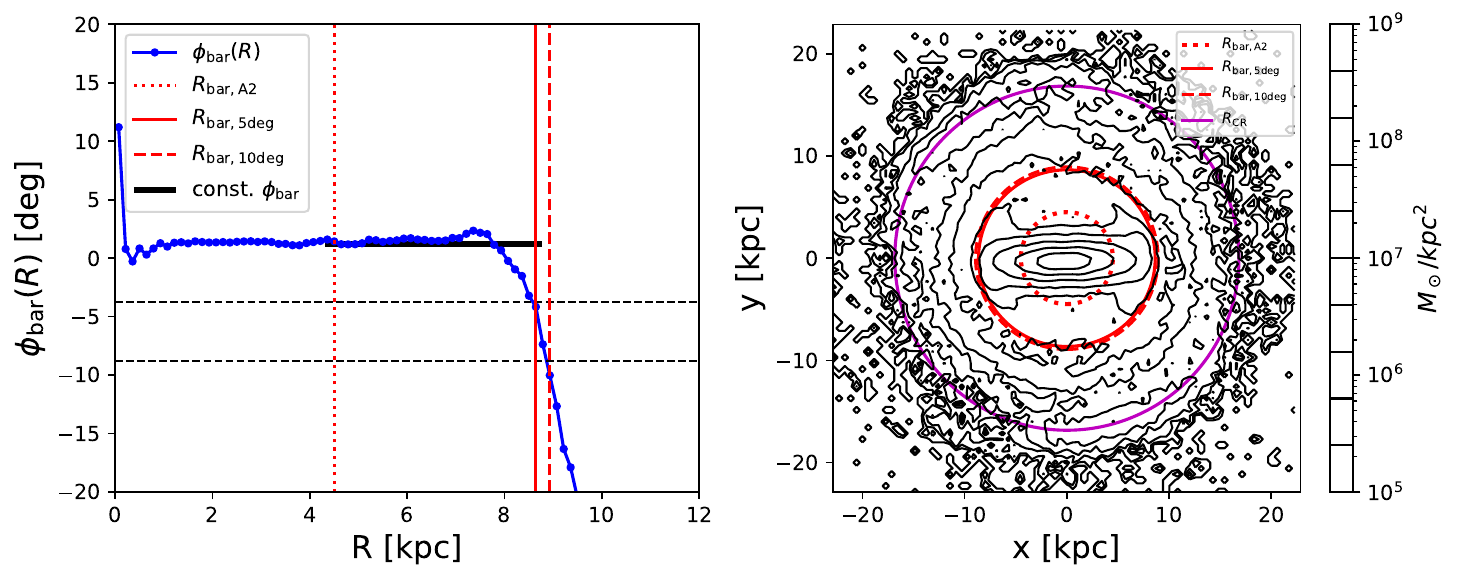}
	\caption{
        Measurement of the length for a stable bar.
        {\it Left panel}: The position angle $\phi_{\mathrm{bar}}(R)$ of the $m=2$ mode of the stellar disk at each radius. The solid black line indicates the averaged position angle of the outer half of the bar, with dashed lines representing deviations of 5$^\circ$ and 10$^\circ$ from the average. The red lines denote the bar length $R_{\mathrm{bar}}$, defined as the radius where $\phi_{\mathrm{bar}}(R)$ deviates from the average by 5$^\circ$ (solid line) and 10$^\circ$ (dashed line).
        The red dotted line represents the bar length $R_{\mathrm{bar, A2}}$, which is the radius at which the bar strength $A_2(R)$ attains its peak.
        {\it Right panel}: The density contour with the bar lengths plotted in red. The line styles are consistent with those in the left panel. The co-rotation radius $R_{\mathrm{CR}}$ is marked with a magenta circle for reference.
    }
	\label{fig:Rbar_stable}
\end{figure}

\begin{figure}
	\includegraphics[width=1\columnwidth]{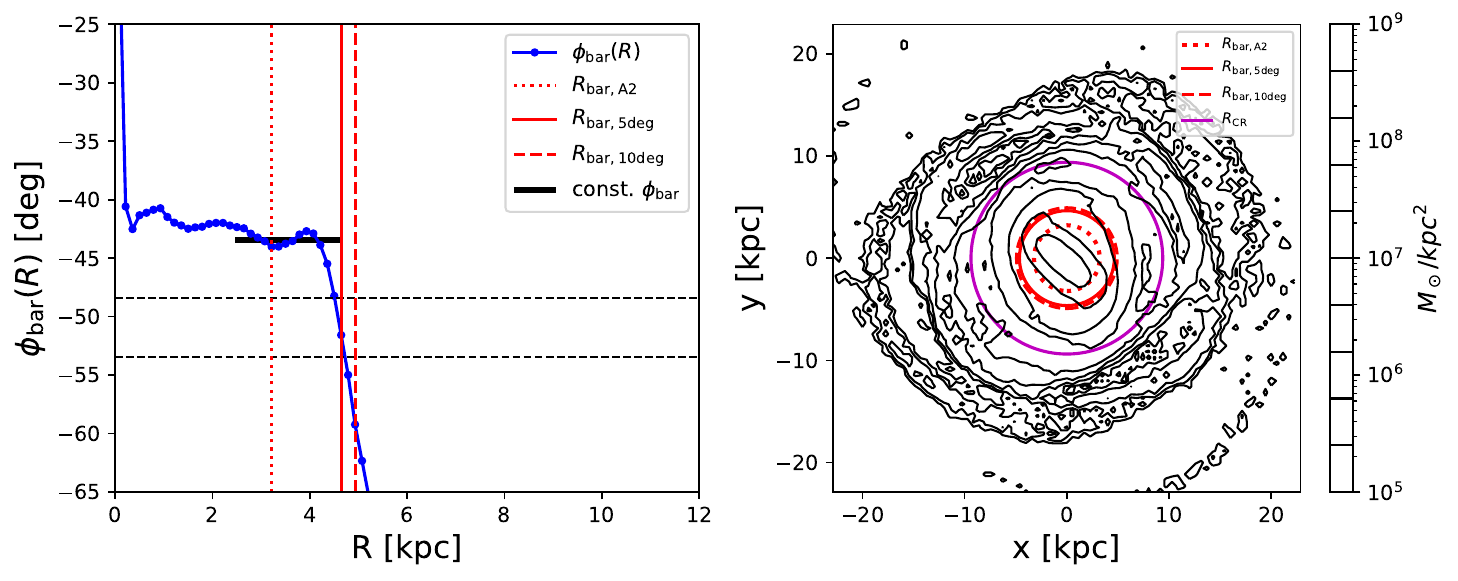}
	\caption{Same as \autoref{fig:Rbar_stable}, but for a bar in the formation stage. The bar is distorted and the $\phi_{\mathrm{bar}}(R)$ varies in the inner and outer parts of the bar. It is a more appropriate choice to use the outer part of the bar as reference (the solid black line) to measure the bar than using the averaged position angle across the entire bar length in this case.}
	\label{fig:Rbar_form}
\end{figure}

To measure the bar length, we evenly sample the radial range with a spacing of 0.15\kpc. 
At each radius, we calculate the phase angle $\phi_{\mathrm{bar}}(R)$ of the $m=2$ mode using stars that fall between ($R-0.3$\kpc, $R+0.3$\kpc). 
The broader width of the annuli is employed to reduce fluctuations in $\phi_{\mathrm{bar}}(R)$.
In the left panel of \autoref{fig:Rbar_stable}, we plot $\phi_{\mathrm{bar}}(R)$ at each radius for a barred disk. 
The bar length is defined as the radius where $\phi_{\mathrm{bar}}(R)$ deviates from a nearly constant value by more than 5$^\circ$. 
The initial estimate for this constant value is the average phase angle within the range of 4\;\Rd. 
We then refine the bar length iteratively by adjusting the reference value to be the average phase angle of the ``outer half'' of the bar. This choice is motivated by two reasons:
1) Not all bars extend to 4\;\Rd, so a fixed range of 4\;\Rd\ may not be suitable for all bars;
2) Bars can be distorted, especially in the formation and buckling stage, and the phase angle may differ in the inner and outer parts of the bar.
An example of a bar in the formation stage is shown in \autoref{fig:Rbar_form} to demonstrate the importance of using the outer part of the bar as a reference for measuring the bar length.
We also plotted the bar length defined by the peak of the bar strength $A_2(R)$ in the left panels of \autoref{fig:Rbar_stable} and \autoref{fig:Rbar_form} for comparison.

In the right panels of \autoref{fig:Rbar_stable} and \autoref{fig:Rbar_form}, we present the density contour with the bar lengths plotted in red. 
The co-rotation radius $R_{\mathrm{CR}}$ is marked with a magenta circle for reference.
The bar length defined by the phase angle variation is consistent with the visually identified bar length in the density contour, whereas the bar length defined by the peak of the bar strength tends to provide underestimated values. 
Increasing the tolerance from 5$^\circ$ to 10$^\circ$ slightly increases the bar length, but the difference is not substantial.

\bibliography{tidal_bar.bib}
\bibliographystyle{aasjournal}

\end{document}